\documentclass[aps,prb,twocolumn,amsmath,amssymb,showpacs,showkeys,floatfix,superscriptaddress]{revtex4-2}
\usepackage{graphicx}
\usepackage{amsfonts}
\usepackage{epsfig}
\usepackage{dcolumn}
\usepackage{amsthm}
\usepackage{color, xcolor}
\usepackage{physics}
\usepackage{latexsym}
\usepackage{amscd}
\usepackage{relsize}
\usepackage{import}
\usepackage{cleveref}

\makeatletter

\begin{document}


\title{Determining the Number of Symmetry Sectors in Composite Quantum Spectra}

\author{Nisarg Vyas}
\email{nisarg.vyas@students.iiserpune.ac.in}
\affiliation{Department of Physics, Indian Institute of Science Education and Research, Pune 411008, India.}

\author{S. Harshini Tekur}
\email{harshini@mindbrug.com}
\affiliation{MindBrug Research LLP, Bengaluru 560085, India}

\author{M. S. Santhanam}  
\email{santh@iiserpune.ac.in}
\affiliation{Department of Physics, Indian Institute of Science Education and Research, Pune 411008, India.}
\date{\today}

\begin{abstract}
The spectral fluctuations in complex quantum systems are modeled in terms of spectra from random matrices. Quantitative agreement with random matrices holds only when the quantum spectrum is a pure sequence, and not a composite spectrum that mixes levels from different symmetry sectors. It was shown earlier that discrete symmetries in composite spectra can be detected using higher-order spacing ratio statistics provided the symmetry sectors are of equal dimensions. The general case is when the symmetry sectors have unequal dimensions, and the symmetries of the Hamiltonian system are unknown or only partially known. In this work, we address how the number of symmetry sectors can be determined from a composite spectrum arising in Hermitian and non-Hermitian settings. A {\it Best Subset Technique} is proposed for determining the number of the symmetry sectors by comparing the spectral fluctuations to that of an appropriate random matrix ensemble without requiring de-symmetrization. It is demonstrated using spectra from Gaussian, Wishart and Ginibre ensembles, and also for spectra drawn from many-body quantum systems. 
 \end{abstract}

\maketitle

{\textit{Introduction:}} Spectral fluctuations in few-body and many-body quantum systems are characterized using the framework of random matrix theory (RMT) \cite{RMT-rev-Guhr1998,akemann2011,Nandkishore_Huse_2015, MBQC-ChanLucaChal2018, MBQC-KosLjuPro2018}. The Bohigas--Giannoni--Schmidt (BGS) conjecture states that quantum systems with a classically chaotic limit exhibit spectral fluctuations consistent with appropriate random matrix ensembles \cite{BGS1984}. In contrast, integrable systems show uncorrelated, typically Poisson-like spectral statistics \cite{Berry_Tabor_1977}. These results were originally formulated for Hermitian Hamiltonians with real spectra belonging to the Wigner–Dyson symmetry classes \cite{mehta2004random}. For non-Hermitian systems with complex spectra, in general, the spectral correlations are consistent with that of the Ginibre ensembles\cite{Ginibre_1965, PhysRevLett.61.1899, PhysRevLett.62.2893}. This has been amply supported by numerical studies of complex spectrum of Liouvillian superoperator \cite{PhysRevLett.123.234103, PhysRevLett.124.100604, PhysRevA.105.L050201, PhysRevResearch.7.013276, Sa_Ribeiro_Prosen_2020}, non-Hermitian effective Hamiltonian \cite{PhysRevLett.123.090603, PhysRevB.106.134202, PRR_Ryuichi, PhysRevResearch.7.013098}, or Lindblad operator \cite{Can_2019, PhysRevLett.123.140403}. Measures such as level spacing distributions \cite{Stockmann_1999, Srivastava_Lakshminarayan_Tomsovic_Backer_2019, PhysRevB.102.054202, Tian_Riser_Kanzieper_2024}, level spacing ratios \cite{oganesyan2007localization, Atas_2013,PhysRevE.101.022222,complex_prosen, Tekur_Bhosale_Santhanam_2018}, as well as the spectral form factor and its variants \cite{Spectral_random,PhysRevLett.127.170602, PhysRevLett.121.264101, chenu_svd} therefore serve as useful diagnostics of chaos. These level fluctuation analysis rely on a crucial prerequisite: the spectrum being analyzed must belong to a single irreducible symmetry sector of the system.

\begin{figure}[t]
\includegraphics[width=0.95\linewidth]{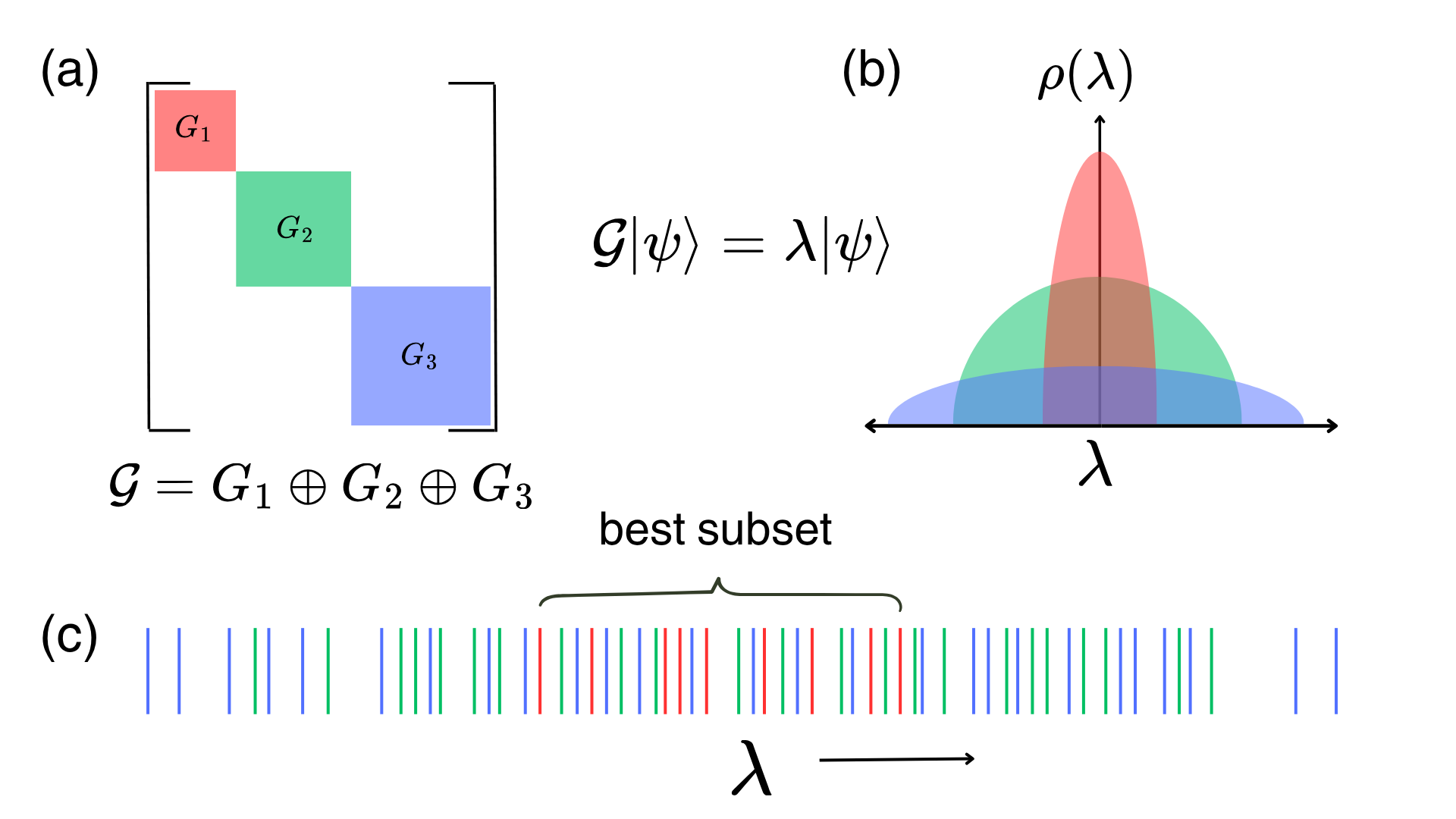}
    \caption{(a) Schematic for a general block diagonal matrix $\mathcal{G}_3 = G_1(N_1) \oplus G_2(N_2) \oplus G_3(N_3)$ with GOE matrices as symmetry sectors. (b) Corresponding symmetry-resolved density of states $\rho(\lambda)$ -- Wigner semicircles of different radii. (c) Composite spectra with colors denoting symmetry sectors. The overlap motivates a technique to deduce the number of blocks.}\label{fig:scaling}
\end{figure}

Discrete symmetries (such as parity, reflection etc.) decompose the Hilbert space into invariant subspaces. A Hamiltonian matrix with $m$ independent symmetry sectors has a block-diagonal structure, and its spectrum is a superposition of $m$ statistically independent level sequences, as illustrated in Fig.~\ref{fig:scaling}(a) for $m=3$. 
Such a composite spectrum does not preserve the characteristic spectral correlations that might exist in the spectra from individual blocks \cite{Berry_Robnik_1984}.
Each block might individually display RMT-type spectral correlations, but the composite spectrum might show Poisson-type statistics. Hence, desymmetrization is essential and can be carried out using symmetry-adapted bases. However, desymmetrization is not straightforward for levels drawn from experiments or complex many-body systems with hidden or only partially known symmetries. Spectral analysis becomes ambiguous in such settings.

The central object of interest is the fluctuations of the composite spectrum $\{E_i\}$ of a Hamiltonian system $H$ disregarding its symmetry structure. Consider a system $H$ with $m$ symmetry blocks, such that their dimensions are all not identical. Given an arbitrary \textit{ordered} sequence of $\{E_i\}$, we address the problem of determining $m$ and their true spectral statistics (say, Wigner or Poisson) without prior knowledge of the symmetry structure or even the system's Hamiltonian. Earlier approaches for this problem suffer from several shortcomings. One approach proposed in \cite{mehta2004random} requires unfolding to be performed, which implies that the level density must be known apriori. Another one constructs a Rosenweig-Porter type surmise for the probability of two consecutive spacings of a composite spectrum \cite{Giraud_Mace_Vernier_Alet_2022}, and an expression for nearest-neighbour spacing ratios are obtained for composite spectrum with $m$ symmetry blocks. However, even for modest values of $m=3,4$, it is almost impossible to distinguish between these cases (see Fig.~2 and Table II in Ref. \cite{Giraud_Mace_Vernier_Alet_2022}). Another approach employs higher order spacing ratios (HOSRs) \cite{Tekur_Bhosale_Santhanam_2018}. It is based on an elegant scaling law for HOSR distributions \cite{TekurPRR2020}, but applies only to matrices drawn from the Gaussian orthogonal ensemble (GOE) with symmetry blocks of equal dimensions, and does not extend to unequal block dimensions or other ensembles. 

In this work, we propose an algorithm -- Best Subset Technique (BST) -- using HOSRs to determine the number of symmetry sectors even when the dimensions of each symmetry block are not identical. We demonstrate it for Gaussian, Wishart-Laguerre, and Ginibre ensembles, as well as for the transverse-field Ising model and a one-dimensional hardcore boson model at half-filling. These are standard RMT ensembles (see Supplementary Material \cite{supp} for their descriptions). The eigenvalues from Gaussian ensembles and Wishart ensembles are real numbers, while the eigenvalues for Ginibre ensemble are complex. Hence, in line with recent results \cite{PhysRevB.111.L161408, PhysRevD.111.L101904}, singular values are analyzed for the Ginibre ensembles\cite{supp}. In the rest of this Letter, $\lambda$ indicates eigenvalues, and $\sigma$ indicates singular values, and $E$ whenever the discussion applies to both $\lambda$ and $\sigma$. 

For an ordered list of eigen- or singular- values 
\begin{equation}
E_1 \le E_2 \le E_3 \dots E_{N-1} \le E_N.     
\end{equation}
drawn from one symmetry sector, the non-overlapping $k$-th order spacing ratio is $r_i^{(k)}=\frac{E_{i+2k}-E_{i+k}}{E_{i+k}-E_i}$, with $i=1,2,\dots,N-2k$. The nearest neighbor $(k=1)$ spacing ratio distribution (SRD) calculated on eigenvalues, has a closed form \cite{Atas_2013}:
\begin{equation}
\label{eq:HOSR-dist}
P(r,\beta)=C_{\beta}\frac{(r+r^2)^{\beta}}{(1+r+r^2)^{(1+\frac{3}{2}\beta)}},
\end{equation}
where $C_{\beta}$ is a normalization constant, and $\beta=1,2,4$ is the Dyson index for the orthogonal, unitary, and symplectic classes, respectively. Eq.~\ref{eq:HOSR-dist} is also a good approximation for the SRD of singular values of Wishart and Ginibre matrices. 

Let $G(N)$ be a GOE matrix of order $N$, and $\{\lambda_i\}$ the $N$ eigenlevels. Consider the direct sum of $m$ GOE matrices 
\begin{equation}
\mathcal{G}^{({\mathbf N})}_m = G_1(N_1)\oplus G_2(N_2) \dots \oplus G_m(N_m)  
\label{eq:dsum1}
\end{equation}
where ${\mathbf N}=(N_1, N_2, \dots, N_m)$ represents the order of the matrices. To fix our ideas, consider the example of $m=3$ GOE matrices with same dimensions, {\it i.e.}, $N_1=N_2=N_3=N$. For the matrices drawn from Gaussian ensembles, eigenvalue support is 
\begin{equation}
\label{eq:scaling_gaussian}
   \Delta_N = 2\sqrt{\beta N}, \;\;\; \text{since } \;\;\; \lambda \in [-\sqrt{\beta N},\sqrt{\beta N}]. 
\end{equation}
Then, $\Delta_N$ of all the 3 GOE matrices coincide. 
Since spectra from each matrix contributes independently to the composite spectrum of $\mathcal{G}^{({\mathbf N})}_3$, the levels mix such that, {\it on average}, there is one level from each matrix in every 3 consecutive levels. Motivated by this interleaving pattern of levels in the composite spectrum, it was shown that for eigenvalues from $\mathcal{G}^{({\mathbf N})}_m$ with $N_1=N_2=\dots=N_m=N$, the $k$-th order spacing ratio distribution ($k$-SRD) is \cite{TekurPRR2020}:
\begin{equation}\label{eq:equal_blocks}
P^k(r,\beta=1,m)=P(r,\beta'), \qquad \beta'=k=m\in\mathbb{Z}.
\end{equation}

Thus, in the ordered spectrum of $\mathcal{G}^{({\mathbf N})}_m$, every $(m+1)$-th eigenvalue is mostly likely to come from the same block. This will be referred to as the \textit{loose interleaving property}. In essence, Eq.~\ref{eq:equal_blocks} suggests that HOSRs can be used as a proxy for the loose interleaving property as long as the participant matrices are of equal sizes. This allows determination of $m$ as that $k$ which satisfies Eq.~\ref{eq:equal_blocks} when $k$-SRDs are calculated over the composite spectrum.

\begin{figure*}
\includegraphics[width=0.99\linewidth, height=0.3\linewidth]{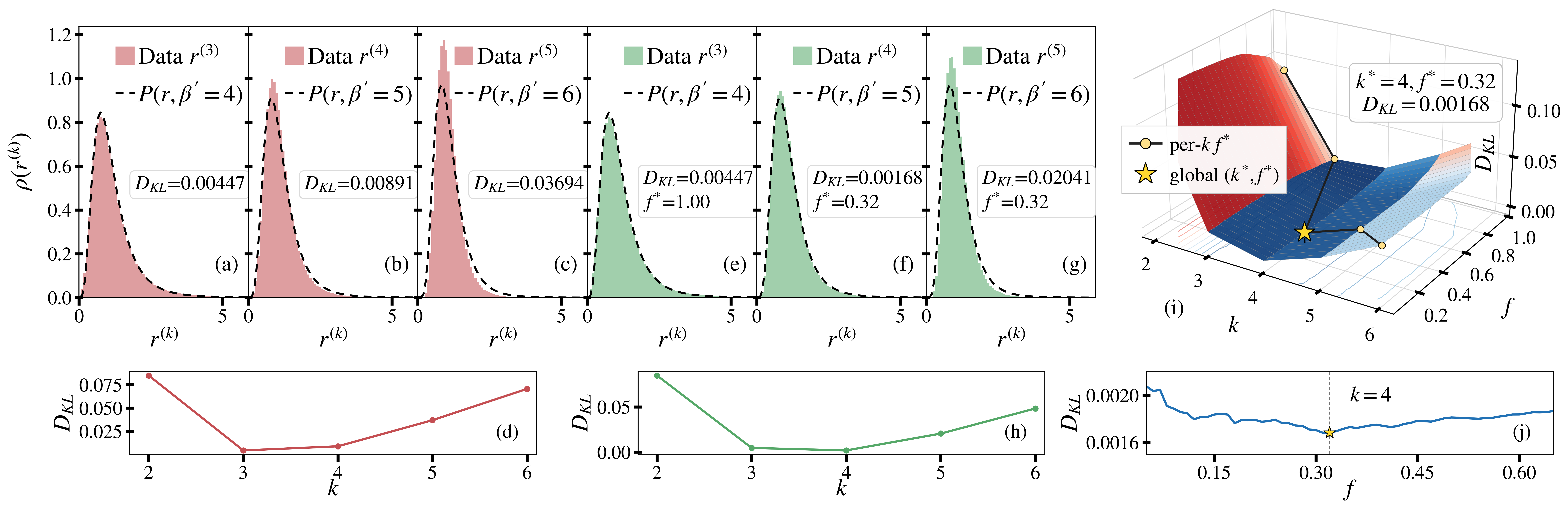}\caption{HOSRs for $k = 2$ to $6$ calculated on composite spectrum of four GUE matrices of sizes $N \in \{5000, 10000, 15000, 20000\}$. In (a-c) we plot the $k$-SRDs with $k=3,4,5$ calculated on the entire spectrum. The dashed line shows Eq.~\ref{eq:HOSR-dist} with $\beta^{\prime} = k+1 = m+1$. (d) $D_{KL}$ minima incorrectly corresponds to $k=3$. (e-g) $k$-SRDs with $k=3,4,5$ with BST applied. (h) Correct deduction of $m = k^*=4$. (i) $D_{KL}(k,f)$ surface as a function of the two parameters of BST, $k$ and $f$. Here the circles show local minima in $f$ for each $k$, and the star locates the global minima in the $k-f$ space. (j) A slice of $D_{KL}(k,f)$ at $k=4$.}
\label{fig:tech-demo-gue}
\end{figure*} 

Now, consider the case $\mathcal{G}^{({\mathbf N})}_3$ with $N_1 < N_2 < N_3$, schematically shown in Fig.~\ref{fig:scaling}(a). The eigenvalue support scales as $\Delta_{N_i} \propto \sqrt{N_i}, i=1,2,3$, which is evident in the level density $\rho(\lambda)$ shown in Fig.~\ref{fig:scaling}(b). 
In this case, $\Delta_{N_i}$ for eigenvalues from the three blocks do not fully overlap. The maximal overlap is near the center of the spectrum, as illustrated in Fig.~\ref{fig:scaling}(c). Hence, the analysis can be restricted to spectral window where eigenvalues from {\it all} blocks overlap. Even then, the loose interleaving property is not guaranteed, since the eigenvalues from $i$-th block has mean spacing $\propto 1/\sqrt{N_i}$. Therefore, for the general case of unequal-size GOE blocks, determination of $m$ using the spectral region of maximal overlap is insufficient due to this size-dependence.

The size-dependence of $\Delta_{N}$ and mean spacing is also present in random matrices from Wishart-Laguerre and Ginibre ensembles. If $X$ is a real/complex matrix of order $M \times N$ ($M\leq N$), such that its columns are standard normal random vectors, then Wishart matrix is defined as $W=XX^{\rm \dagger}$, and the eigenvalue support is 
\begin{equation}
\label{eq:scaling_wishart}
\Delta_N = 4\sqrt{M N}, \;\; \text{since} \;\;  \lambda \in [N(1-c)^2, N(1+c)^2],
\end{equation}
where $c=\sqrt{M/N}$. In the special case $N=M$, asymptotically, $\lambda \in (0,4N)$. Hence for $m$ unequal-size blocks of Wishart matrices, each with $M_i=N_i$, the region of maximal overlap lies at the lower end of the composite spectrum. If $Y$ is an order $N$ Ginibre random matrix, then singular values of $Y$ are the eigenvalues of $\sqrt{YY^{\dagger}}$, and the singular-value support is
\begin{equation}
\label{eq:scaling_ginibre}
\Delta_N = 2\sqrt{N}, \;\; \text{since} \;\;  \sigma \in [0, 2\sqrt{N}].
\end{equation}
Once again, this scaling implies that for $m$ unnormalized Ginibre matrices of different sizes, the region of maximal overlap lies at the lower end of the composite spectrum. Hence, Eq.~\ref{eq:scaling_gaussian}, \ref{eq:scaling_wishart} and \ref{eq:scaling_ginibre} together imply that a fraction $f$ of eigen/singular values picked from the center (lower end) of the composite spectrum of Gaussian (Wishart/Ginibre) matrices carries more information about all the contributing blocks than arbitrary subsets picked from elsewhere in the composite spectrum. This suggests a general technique that treats both $f$ and $k$ as free parameters and determines $m$. The central assumption is that HOSRs can still be a proxy for the loose interleaving property if the eigenvalues are drawn from a suitable subset of the composite spectrum. In its general form, this does not appear to be an analytically tractable problem. Hence we propose a numerical technique described as follows.

\textit{Best subset technique}:  

The best fraction, $f^*$, and the best order $k^*$ are computed via an iterative procedure based on the central result of this paper:
\begin{equation}
\label{eq:unequal_blocks}
P^k(r(f),\beta,m)= P(r,\beta'), 
\end{equation}
where $r(f)$ are the HOSRs on a fraction $f$ of the ordered levels in the composite spectrum. Then $k^*$ and $f^*$ are the optimal values such that Eq.~\ref{eq:unequal_blocks} holds. Inference of $m$ is then self-consistently made via the following relations
\begin{equation}
\begin{split}
    \beta'=k=m \;, & \qquad \text{for GOE, LOE, GinOE}, \\
    \beta'=k+1=m+1 \;, & \qquad \text{for GUE, LUE, GinUE}, \\
    \beta'=k+3=m+3 \;, & \qquad \text{for GSE, LSE, GinSE}.
\end{split}
\label{eq:unequal_blocks1}
\end{equation}
Note that Eqs.~\ref{eq:unequal_blocks}-\ref{eq:unequal_blocks1} are a generalization of Eq.~\ref{eq:equal_blocks} beyond the GOE case. The inputs to BST are a maximum order $K_{max} \ll L=\sum_{i=1}^m N_i$, and a sorted list of composite spectrum (eigen- or singular- values) $E = [E_1, \ldots, E_L]$, from $m$ matrices of arbitrary sizes. Iterating over $f \in (0,1)$ and $k \in \{1,2, \dots, K_{max}\}$, BST computes the $k$-SRD over a fraction $f$ of the composite spectrum, and $k^*$ and $f^*$ are defined as
\begin{equation}
    k^*,f^* = \underset{k,f}{\rm argmin} \; D_{\mathrm{KL}}(k,f)
    \label{eq:optimal_vals}
\end{equation}
where $D_{\mathrm{KL}}(k,f)$ is the Kullback–Leibler (KL) divergence between the empirical $k$-SRD computed on a fraction $f$ of the composite spectrum (\textit{lhs} of Eq.~\ref{eq:unequal_blocks}), and the analytical form of Eq.~\ref{eq:HOSR-dist} with $\beta^{\prime} = k$ (\textit{rhs} of Eq.~\ref{eq:unequal_blocks})
$$D_{\mathrm{KL}}(k,f) = \sum_{i=1}^{n-2k} P^{(k)}\left(r_i(f), \beta,m\right) \log\left(\frac{P^{(k)}(r_i(f), \beta,m)}{P(r_i,\beta'=k)}\right).$$

\textit{Random matrices:} Applying the BST on GUE random matrices, Fig.~\ref{fig:tech-demo-gue} shows that the number of blocks $m$ can be deduced via HOSRs calculated on the fraction $f^*$ of composite spectrum. Without BST, as shown in Fig.~\ref{fig:tech-demo-gue}(a-c), the minima of $D_{KL}$ incorrectly identifies $\beta'=4$ and hence $k^*=3$ as seen in Fig.~\ref{fig:tech-demo-gue}(d). Evidently, HOSRs drawn from the composite spectrum as a proxy for the loose interleaving property incorrectly implies $m=3$ blocks. In contrast, applying BST in Fig.~\ref{fig:tech-demo-gue}(e-g), $D_{KL}$ correctly identifies $\beta'=5$ and hence $k^*=4$ as seen in Fig.~\ref{fig:tech-demo-gue}(h). Indeed, Fig.~\ref{eq:HOSR-dist}(i) illustrates that the identified pair $(k^*=4,f^* \approx 32\%)$, based on Eqs. \ref{eq:unequal_blocks}-\ref{eq:optimal_vals}, corresponds to the global minima of $D_{KL}$ (see Fig.~\ref{eq:HOSR-dist}(j) for a clearer view of minima in $f$). Thus, HOSRs restricted to the best fraction $f^*$ carries a better imprint of loose interleaving property. Plots similar to Fig.~\ref{eq:HOSR-dist} for all the other RMT ensembles are shown in \cite{supp}.

\begin{figure}[t]
\includegraphics[width=0.99\linewidth]{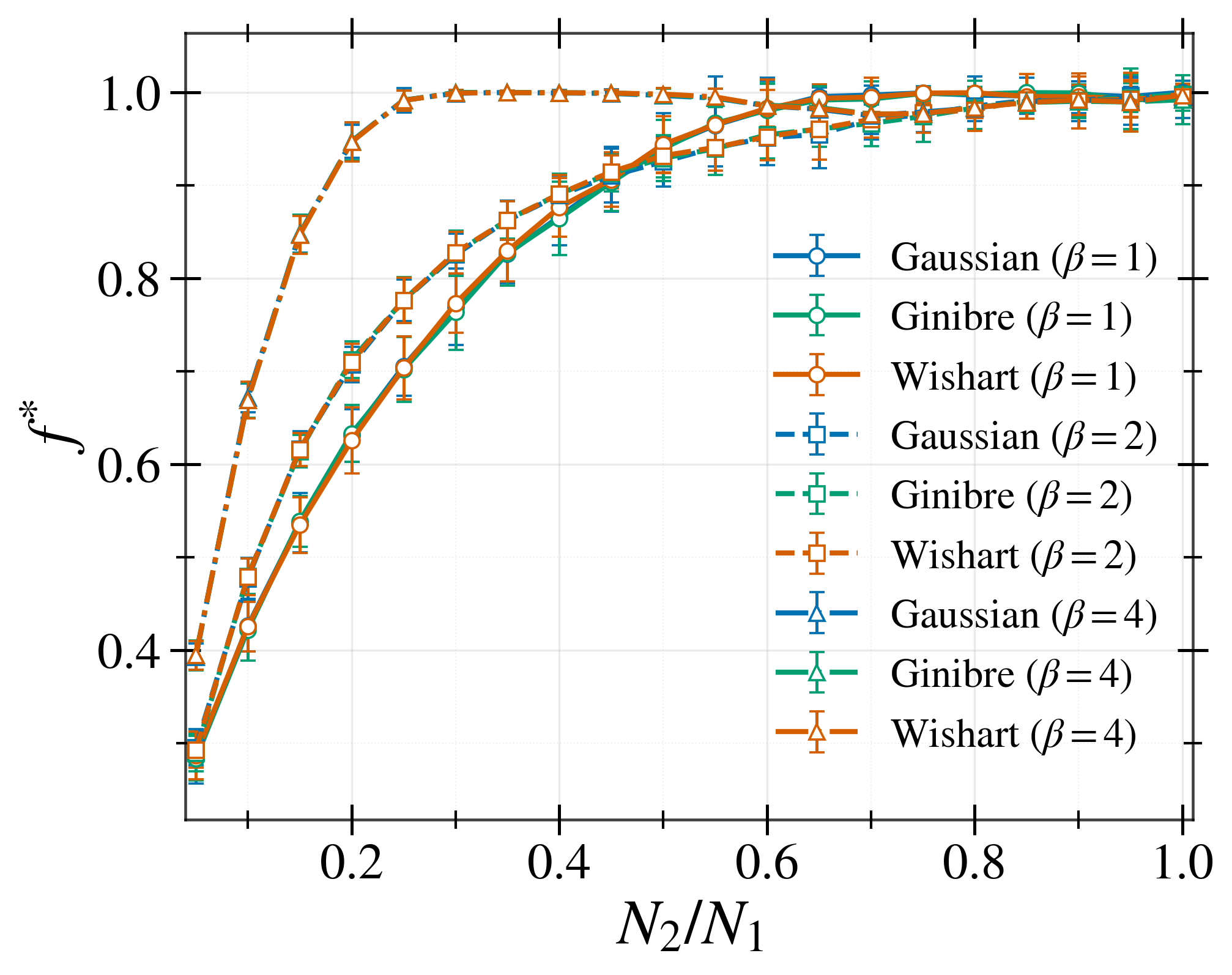}\caption{The best fraction $f^*$ for composite spectra from two random matrices of orders $N_1$ (fixed at $20000$) and $N_2$ (varying). All the $\beta=1$ ensembles fall onto one curve (circles). Similarly, all $\beta=2$ and $\beta=4$ ensembles follow the curves denoted by squares and triangles respectively.}
\label{fig:fraction_vs_ratio}
\end{figure}

A particularly instructive case is that of two independent blocks, $m=2$ with dimensions $N_1$ and $N_2$. 
Figure~\ref{fig:fraction_vs_ratio} shows a non-trivial scaling of $f^*$ with the ratio $N_2/N_1$. Note that $f^* \to 1$ as $N_2/N_1 \to 1$. Even for block-size ratios $\sim 0.6$, $f^* \gtrsim 0.95$. Remarkably, $f^*$ is independent of the ensemble but depends on the symmetry class indexed by the Dyson index $\beta$. The data from Gaussian (blue), Ginibre (green), and Wishart (red) ensembles with the same value of $\beta$ (say, $\beta=1$) lie on the same curve. Thus, the best fraction $f^*$ is governed primarily by $\beta$ and the relative block size $N_2/N_1$ rather than microscopic ensemble details. This merit a more detailed investigation, that will be addressed in a separate work. Note that this collapse may be attributed partially to the validity of Eq.~\ref{eq:HOSR-dist} for all the ensembles under consideration. This guarantees that their empirical $k$-SRD are always compared against the same analytical form.

\begin{figure}[t]
    \includegraphics[width=0.99\linewidth]{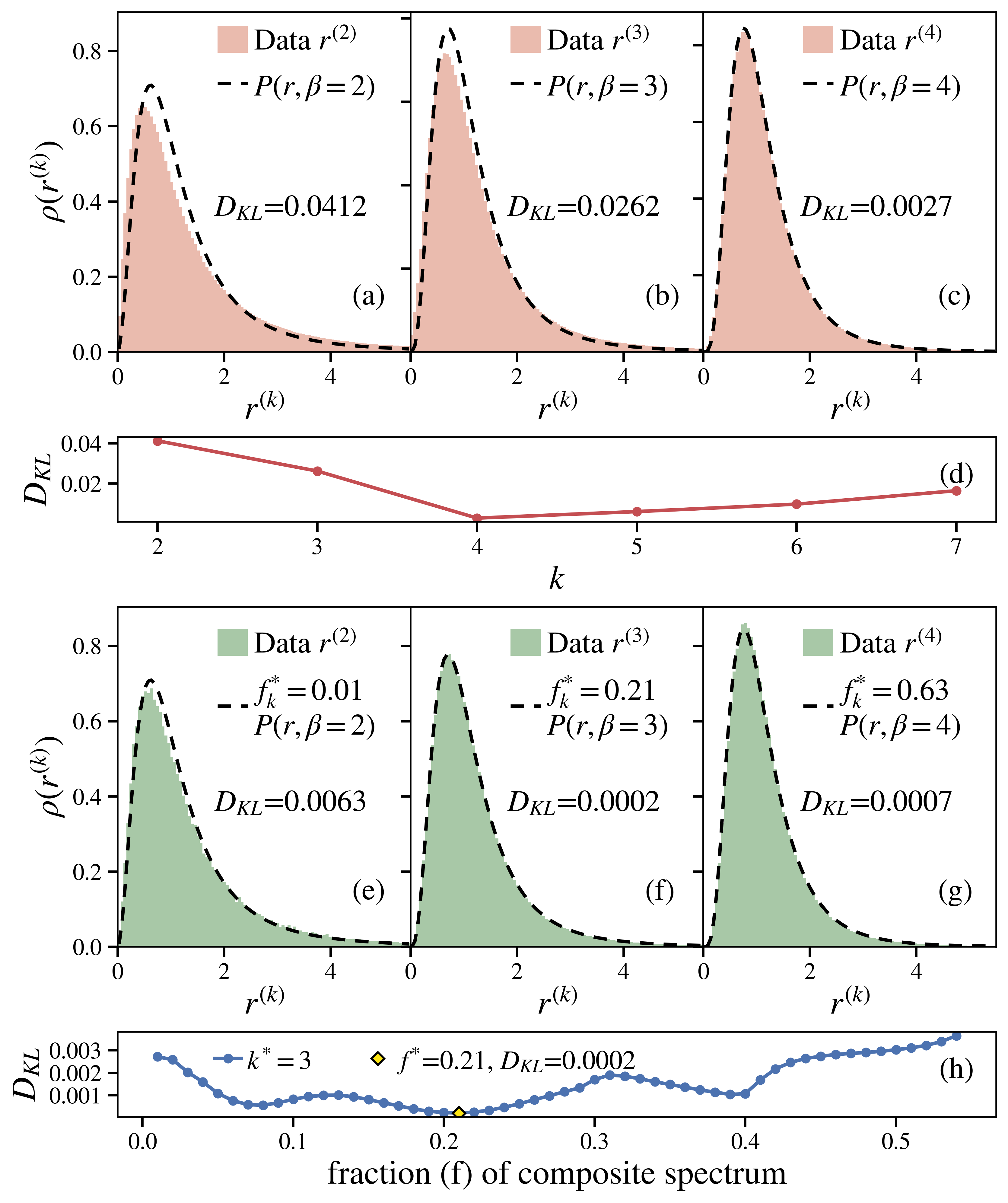}
    \caption{HOSRs calculated on composite Hamiltonian spectrum from three TFIMs with 14, 15 and 16 quantum spins. (a-c) HOSRs for $k = 2,3,4$ obtained from the composite spectrum. (d) Incorrect deduction that $m=4$. (e-g) HOSRs with BST applied. (h) shows the minima in $f$ when $k=3$.}
    \label{fig:tfi}
\end{figure}

\textit{Transverse Field Ising Model:}
To illustrate BST on a physical system, consider a disordered, all-to-all transverse field Ising model (TFIM), whose Hamiltonian is
\begin{equation}
    H = \sum_{i < j}^N J_{ij} \sigma^z_i \sigma^z_j + \sum_{i=1}^N h_i \sigma^x_i.
\end{equation}
In this, $\sigma^z_i$ and $\sigma^x_i$, respectively, are the longitudinal and transverse Pauli operators for the $i$-th spin. The longitudinal couplings $J_{ij}$ are sampled from a Gaussian distribution $J_{ij} \sim \mathcal{N}(\mu_J, \sigma_J)$ with mean $\mu_J = 5.0$ and standard deviation $\sigma_J = 0.5$. The transverse fields $h_i$, sampled from $h_i \sim \mathcal{N}(\mu_h, \sigma_h)$ with $\mu_h = 1.0$ and $\sigma_h = 0.1$, break any remaining translational or reflection symmetries. Thus the chosen configuration ensures chaotic dynamics for this model \cite{GraSwaHey2025,RussFavaHeyl2021}. The spectral analysis in Fig.~\ref{fig:tfi} is performed on the composite spectrum constructed from systems of sizes $N \in \{14, 15, 16\}$, and the results averaged over 100 independent disorder realizations. Since the correct random matrix ensemble is GOE, we expect the analytical $P(r,\beta^{\prime} = m = 3)$ to match in distribution with the empirical $3$-SRD calculated on the composite spectrum. Fig.~\ref{fig:tfi}(a-d) shows incorrect conclusion that $k^*=m=4$. Then, on applying the BST, we obtain the correct answer $k^*=m=3$ for the case when only $f^* \approx 21\%$ of the ordered composite spectra is used for the HOSR computations. 

\begin{figure}[t]
\includegraphics[width=1.0\linewidth]{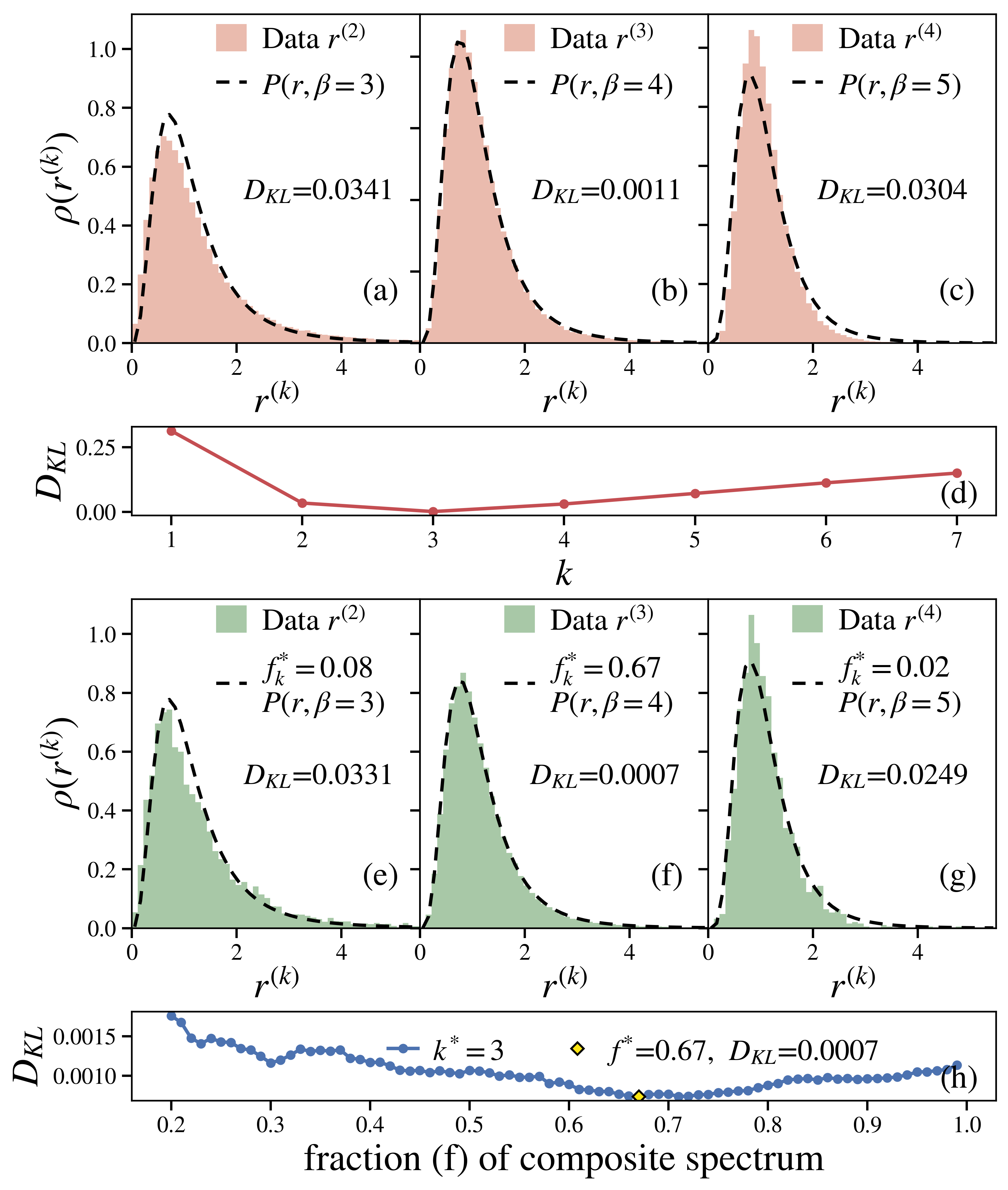}
    \caption{HOSRs calculated on the composite spectrum from three hard-core boson models with $16$, $17$ and $18$ bosons. (a-d) The KL divergence between $P(r,\beta'=k)$ and $P^{(k)}(r, \beta,m)$ reveals $m=3$. (e-h) Application of BST confirms $m=3$.}
    \label{fig:hardcorebosons}
\end{figure}

\textit{Non-Hermitian Hamiltonian:}
To demonstrate BST on non-Hermitian systems, we consider a one-dimensional non-reciprocal hardcore boson model at half-filling \cite{HamaKawaUeda2019}. This model incorporates non-Hermiticity through both asymmetric hopping, and a complex staggered on-site potential. The Hamiltonian is :
\begin{equation}
H = \sum_{j=1}^L \left[ -t \left( e^g b_j^\dagger b_{j+1} + e^{-g} b_{j+1}^\dagger b_j \right) + V n_j n_{j+1} + \mu_j n_j \right],
\end{equation}
where $b_j^\dagger$ ($b_j$) are the creation (annihilation) operators for hardcore bosons satisfying the constraint $b_j^{\dagger 2} = 0$, and $n_j = b_j^\dagger b_j$ is the local density operator. The first term describes non-reciprocal hopping with amplitude $t=2.0$ and a skin-effect parameter $g=0.1$, which breaks Hermiticity by favoring one direction of transport. Interaction is included via a nearest-neighbor term with strength $V=2.0$. The local potential $\mu_j = W_j + i(-1)^j \gamma$ contains a disordered real part $W_j \in [-h, h]$ (with $h=2$), and an imaginary part $\gamma=0.1$ representing alternating gain and loss. We enforce periodic boundary conditions and restrict the Hilbert space to the half-filling sector $N_b = L/2$ for system sizes $L \in \{16, 17, 18\}$. Composite spectra for each system size is obtained for 100 independent disorder realizations. Since this model corresponds to the GinUE class, we expect $P(r,\beta^{\prime} = m+1 = 4)$ to match the $3$-SRD calculated on the singular values. Note that, Fig.~\ref{fig:hardcorebosons}(a-c) correctly show this without the use of the \textit{best-subset technique} which only confirms the finding in Fig.~\ref{fig:hardcorebosons}(e-g). This is because for this particular model the spectrum does not show any scaling with the system size, and hence the overlap between spectra from the non-interacting systems of various sizes is the same as if they were of the same size. 


\textit{Summary:} The key observation is that higher-order spacing ratios are sensitive to the number of symmetry sectors even when the sectors are of unequal dimensions. This work exploits this idea and introduces an empirical approach -- the Best Subset Technique -- to infer the number of symmetry sectors in complex quantum systems using higher order spectral fluctuations without constructing symmetry-adapted bases and spectral unfolding. The BST selects the best subset of levels from a composite spectrum based on the scaling of maximal and minimal eigenvalues with random matrix dimension $N$. This approach is general and applies equally to spectra from Gaussian, Wishart and Ginibre ensembles, the latter applicable to non-Hermitian and open quantum systems. For $m=2$, the best fraction depends only on the Dyson index $\beta$ and not on the details of the random matrix ensemble. BST is most effective when the composite spectrum exhibits a sufficiently large overlap region, i.e, when the number of levels are large for reliable statistics. In future, analytical support for these results would be useful, as also extension to Wishart ensembles with $N\neq M$ for which both the spectral edges scale non-trivially with system size.

{\sl Acknowledgment.}--- NV thanks TCS Foundation for financial support with the TCS RSP Fellowship (2024-2028). MSS would like to acknowledge ARG-MATRICS grant ARGM/2026/0176 from ANRF, Govt of India.

\bibliography{symmViaHOSR} 
\onecolumngrid
\clearpage
\onecolumngrid    
\section*{SUPPLEMENTARY MATERIAL}

\subsection{Description of relevant random matrix ensembles}
The Gaussian random matrix ensembles are defined as families of Gaussian-distributed random matrices, and depending on whether their distributions are invariant under orthogonal, unitary or symplectic transformations, are classified as Gaussian Orthogonal Ensemble(GOE), Gaussian Unitary Ensemble(GUE), and Gaussian Symplectic Ensembles(GSE). These are also classified according to their Dyson index ($\beta$) with $\beta=1,2$ and $4$ for the orthogonal, unitary and symplectic ensembles, respectively. For random matrices of order $N$, drawn from the standard Gaussian ensembles, the analysis will be carried out on the eigenvalues $\{\lambda_l\}_{l=1}^N$. Random matrices $W$ from the Wishart-Laguerre ensemble are defined as follows: let $X$ be $M \times N$ ($M\leq N$) rectangular matrix with elements drawn from $\mathcal{N}(0,1)$,
Gaussian distribution with zero mean and unit variance. Then, $W = XX^{\dagger}$, where $\dagger$ denotes the conjugate transpose for matrices whose entries are complex numbers (for Laguerre Unitary Ensemble (LUE)), or quaternions (for Laguerre Symplectic Ensemble (LSE)), or a standard transpose {\rm T} if the matrix entries are real numbers (for Laguerre Orthogonal Ensemble(LOE)). The quantities of interest are $\lambda_l \geq 0$, the eigenvalues of $W$ . 
The Ginibre ensemble is a collection of random matrices whose entries are drawn from $\mathcal{N}(0,1)$ without any Hermiticity or symmetry constraints. The elements of Ginibre [Orthogonal (GinOE), Unitary (GinUE), Symplectic (GinSE)] ensembles, are respectively, purely real, complex and quaternion numbers. For non-Hermitian Hamiltonians and for Ginibre matrices, the eigenvalues are complex numbers, and we work with singular values $\{\sigma_l\}_{l=0}^{N}$ which are real and defined as the square-root of the eigenvalues of $G^{\dagger}G$. For the Ginibre class, Higher Order Spacing Ratios (HOSRs) of singular values can distinguish between the symmetry classes and capture long-range correlations. 

\subsection{Best Subset Technique illustrated for the HOSRs of the Gaussian, Ginibre and Wishart-Laguerre Ensembles}

\begin{figure*}[h]
\includegraphics[width=0.99\linewidth]{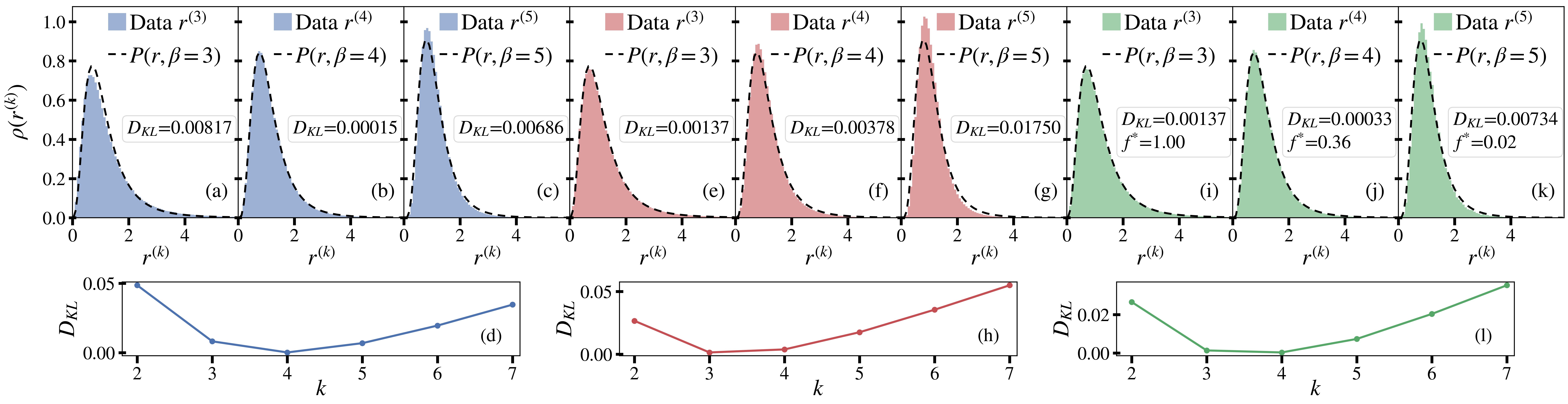}\caption{HOSRs for $k = 2$ to $7$ calculated on composite spectrum of four GOE matrices. (a-c) Application of Eq.~\ref{eq:equal_blocks} to the four equal-size matrices of size $10000$, and (d) correct detection of the number of blocks. In (e-g) we show symmetry deduction via Eq.~\ref{eq:equal_blocks} applied to matrices of sizes $N \in \{5000, 10000, 15000, 20000\}$, which leads to incorrect conclusions as shown in (h). This is fixed in (i-k) by using the best-subset technique which detects the correct number of blocks by using only a fraction of the levels. Note that owing to the scaling in Eq.~\ref{eq:scaling_gaussian}, this fraction is chosen from the center of the ordered composite spectrum.}
\end{figure*}

\begin{figure*}[h]
\includegraphics[width=0.99\linewidth]{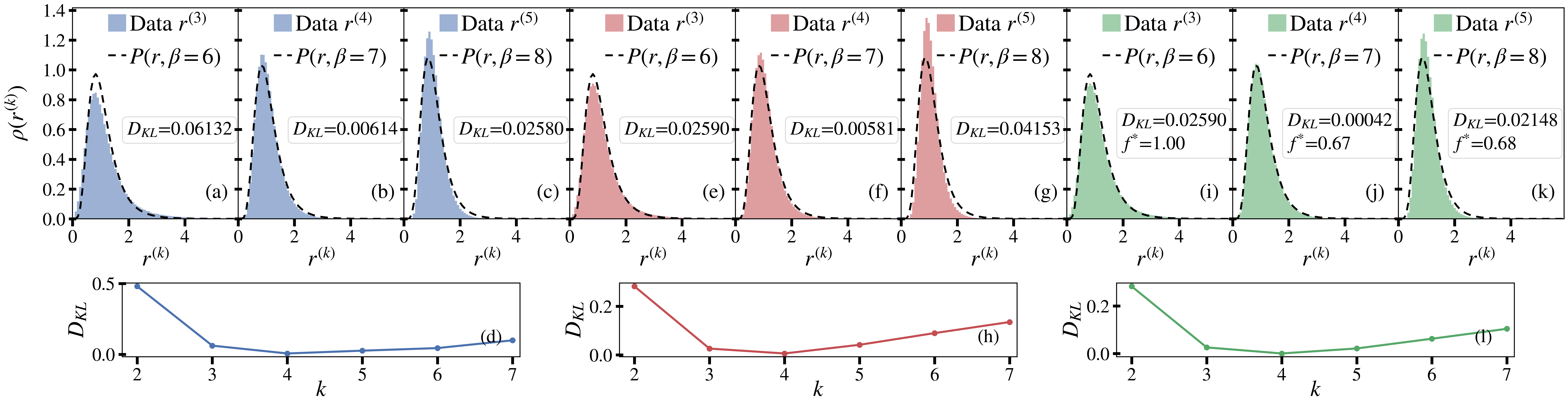}\caption{HOSRs for $k = 2$ to $7$ calculated on composite spectrum of four  matrices sampled from Gaussian Symplectic Ensemble (GSE). (a-c) Application of Eq.~\ref{eq:equal_blocks} to the four equal-size matrices of size $10000$, and (d) correct detection of the number of blocks. In (e-g) we show symmetry deduction via Eq.~\ref{eq:equal_blocks} applied to matrices of sizes $N \in \{5000, 10000, 15000, 20000\}$, which leads to incorrect conclusions as shown in (h). This is fixed in (i-k) by using the best-subset technique which detects the correct number of blocks by using only a fraction of the levels. Note that owing to the scaling in Eq.~\ref{eq:scaling_gaussian}, this fraction is chosen from the center of the ordered composite spectrum.}
\end{figure*}

\begin{figure*}
\includegraphics[width=0.99\linewidth]{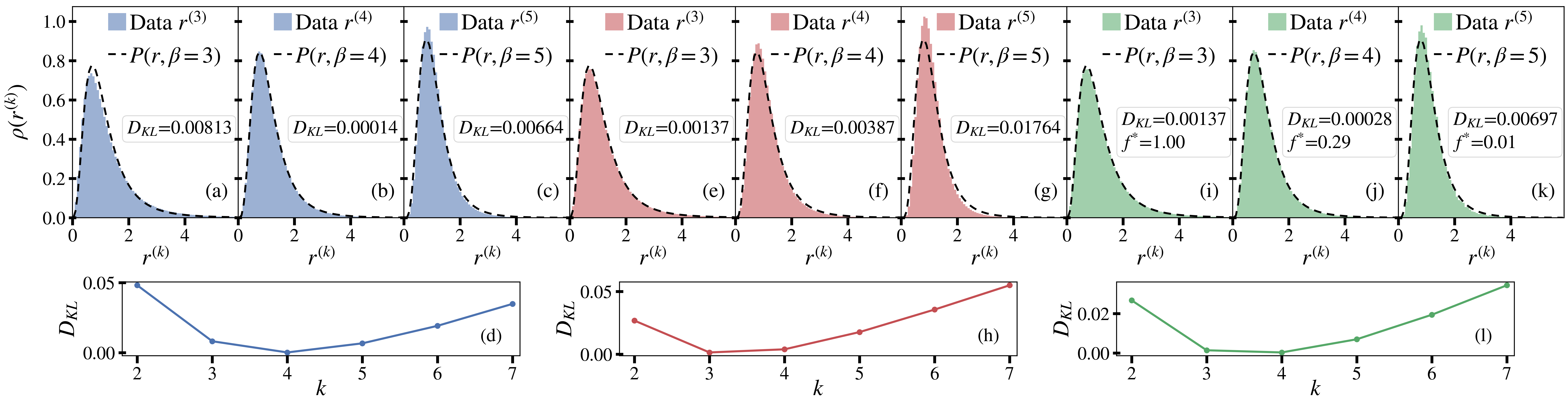}\caption{HOSRs for $k = 2$ to $7$ calculated on the composite singular value spectrum of four GinOE matrices. (a-c) Application of Eq.~\ref{eq:equal_blocks} to the four equal-size matrices of size $10000$, and (d) correct detection of number of blocks. In (e-g) we show symmetry deduction via Eq.~\ref{eq:equal_blocks} applied to matrices of sizes $N \in \{5000, 10000, 15000, 20000\}$, which leads to incorrect conclusions as shown in (h). This is fixed in (i-k) by using the best-subset technique which detects the correct number of blocks by using only a fraction of the levels. Since singular values are non-negative, the overlap of spectra lies in the lower end. Hence, the best fraction is chosen from the lower end of the ordered composite spectrum.}
\end{figure*}

\begin{figure*}
\includegraphics[width=0.99\linewidth]{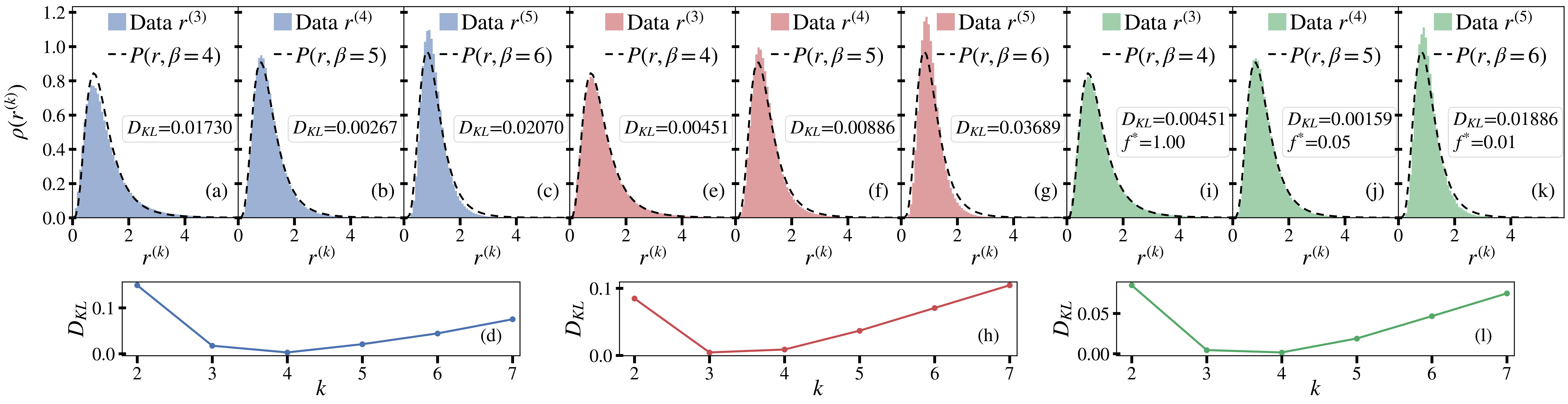}\caption{HOSRs for $k = 2$ to $7$ calculated on the composite singular value spectrum of four GinUE matrices. (a-c) Application of Eq.~\ref{eq:equal_blocks} to the four equal-size matrices of size $10000$, and (d) correct detection of number of blocks. In (e-g) we show symmetry deduction via Eq.~\ref{eq:equal_blocks} applied to matrices of sizes $N \in \{5000, 10000, 15000, 20000\}$, which leads to incorrect conclusions as shown in (h). This is fixed in (i-k) by using the best-subset technique which detects the correct number of blocks by using only a fraction of the levels. The best fraction is chosen from the lower end of the ordered composite spectrum.}
\end{figure*}

\begin{figure*}
\includegraphics[width=0.99\linewidth]{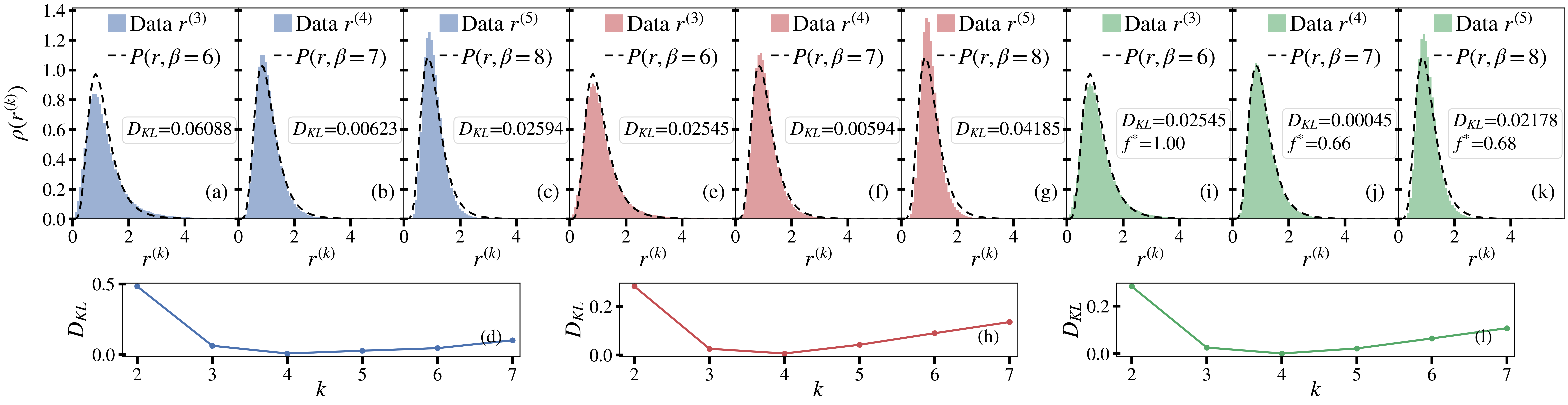}\caption{HOSRs for $k = 2$ to $7$ calculated on the composite singular value spectrum of four GinSE matrices. (a-c) Application of Eq.~\ref{eq:equal_blocks} to the four equal-size matrices of size $10000$, and (d) correct detection of number of blocks. In (e-g) we show symmetry deduction via Eq.~\ref{eq:equal_blocks} applied to matrices of sizes $N \in \{5000, 10000, 15000, 20000\}$, which leads to incorrect conclusions as shown in (h). This is fixed in (i-k) by using the best-subset technique which detects the correct number of blocks by using only a fraction of the levels. Since singular values are non-negative, the overlap of spectra lies in the lower end. Hence, the best fraction is chosen from the lower end of the ordered composite spectrum.}
\end{figure*}

\begin{figure*}
\includegraphics[width=0.99\linewidth]{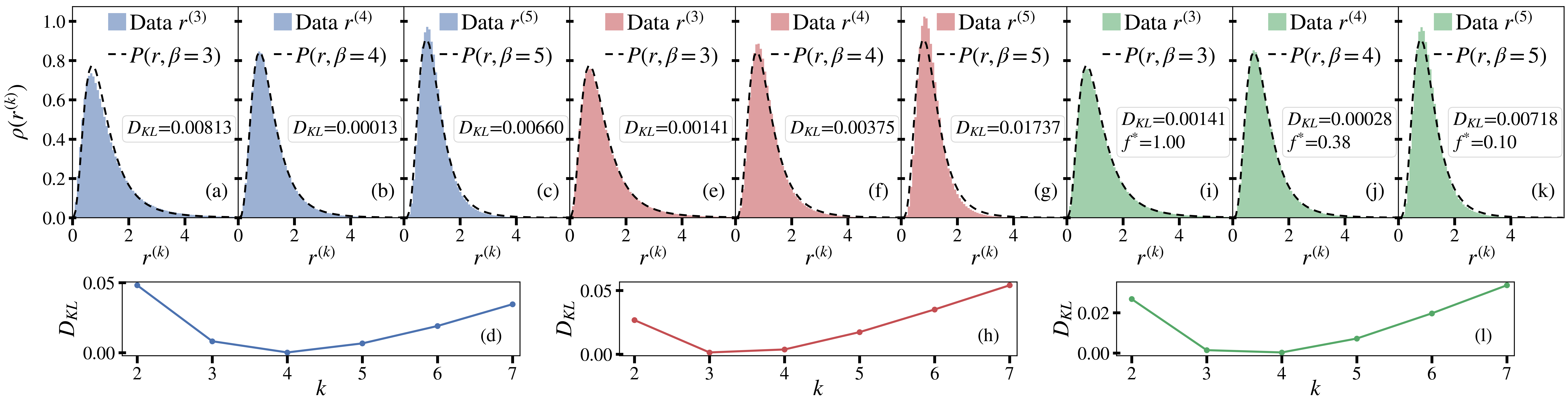}\caption{
HOSRs for $k = 2$ to $7$ calculated on the composite spectrum of four LOE matrices. (a-c) Application of Eq.~\ref{eq:equal_blocks} to the four equal-size matrices of size $10000$, and (d) correct detection of the number of blocks. In (e-g) we show symmetry deduction via Eq.~\ref{eq:equal_blocks} applied to matrices of sizes $N \in \{5000, 10000, 15000, 20000\}$, which leads to incorrect conclusions as shown in (h). This is fixed in (i-k) by using the best-subset technique which detects the correct number of blocks by using only a fraction of the levels. Note that, owing to the scaling of spectral bounds in Eq.~\ref{eq:scaling_wishart}, this fraction is chosen from the lower end of the ordered composite spectrum.
}
\end{figure*}

\begin{figure*}[t]
\includegraphics[width=0.99\linewidth]{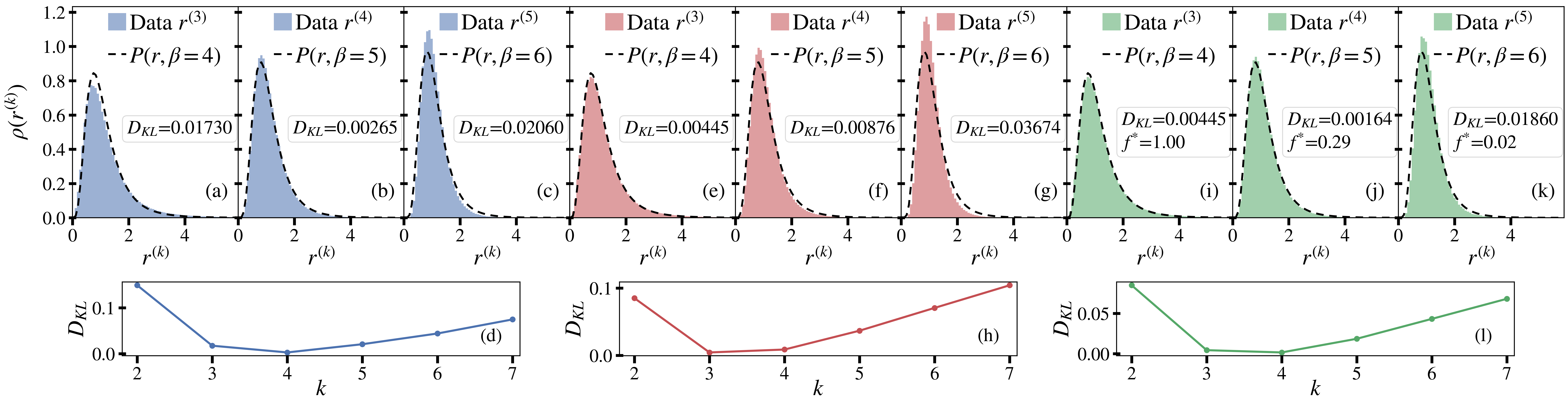}\caption{
HOSRs for $k = 2$ to $7$ calculated on the composite spectrum of four LUE matrices. (a-c) Application of Eq.~\ref{eq:equal_blocks} to the four equal-size matrices of size $10000$, and (d) correct detection of the number of blocks. In (e-g) we show symmetry deduction via Eq.~\ref{eq:equal_blocks} applied to matrices of sizes $N \in \{5000, 10000, 15000, 20000\}$, which leads to incorrect conclusions as shown in (h). This is fixed in (i-k) by using the best-subset technique which detects the correct number of blocks by using only a fraction of the levels. Note that, owing to the scaling of spectral bounds in Eq.~\ref{eq:scaling_wishart}, this fraction is chosen from the lower end of the ordered composite spectrum.
}
\end{figure*}

\begin{figure*}[t]
\includegraphics[width=0.99\linewidth]{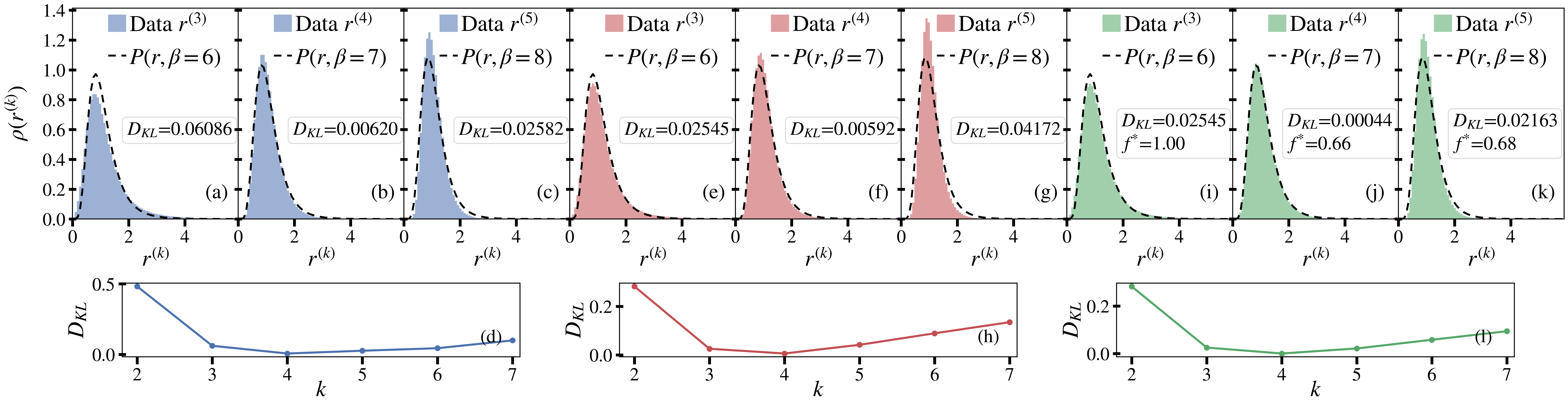}\caption{
HOSRs for $k = 2$ to $7$ calculated on the composite spectrum of four LSE matrices. (a-c) Application of Eq.~\ref{eq:equal_blocks} to the four equal-size matrices of size $10000$, and (d) correct detection of the number of blocks. In (e-g) we show symmetry deduction via Eq.~\ref{eq:equal_blocks} applied to matrices of sizes $N \in \{5000, 10000, 15000, 20000\}$, which leads to incorrect conclusions as shown in (h). This is fixed in (i-k) by using the best-subset technique which detects the correct number of blocks by using only a fraction of the levels. Note that, owing to the scaling of spectral bounds in Eq.~\ref{eq:scaling_wishart}, this fraction is chosen from the lower end of the ordered composite spectrum.
}
\end{figure*}

\end{document}